\documentclass[preprint,12pt]{article}

\usepackage[left=1.0in, top=1.0in, right=1.0in, bottom=1.0in, headsep=0.0in]{geometry}
\usepackage{graphics,graphicx,amssymb,hyperref,setspace,enumitem}
\usepackage{color}
\usepackage{titlesec}

\usepackage[numbers,sort&compress]{natbib}
\usepackage{changepage}

\definecolor{myblue}{rgb}{0.250, 0.250, 0.7500}
\definecolor{myblue}{rgb}{0, 0, 0.5}
\definecolor{myred}{rgb}{0.500, 0.250, 0.250}

\newcommand{\doi}[1]{\url{http://dx.doi.org/#1}}

\newcommand{\tild}{\raise.17ex\hbox{$\scriptstyle\mathtt{\sim}$}}
\newcommand{\blue}[1]{{\color{myblue}#1}}

\newcommand{\msigma}{M--$\sigma^{*}$}

\usepackage{lineno}

\begin{document}

\titleformat*{\section}{\large\bfseries\blue}
\titleformat*{\subsection}{\bfseries\blue}

\titlespacing*{\section}{0pt}{6pt}{1pt}
\titlespacing*{\subsection}{0pt}{6pt}{1pt}
\vspace{-2cm}
\Large
\begin{center}
\textbf{MULTI-MESSENGER ASTROPHYSICS \\ WITH PULSAR TIMING ARRAYS} \\
\textcolor{myblue}{\textbf{Astro2020 Science White Paper}} \\
\end{center}

\normalsize

\begin{flushleft}
\noindent \textcolor{myblue}{\textbf{\large Thematic Areas}}
$\rlap{X}\square$ Multi-Messenger Astronomy \& Astrophysics \linebreak
\hspace*{117pt}$\rlap{X}\square$ Galaxy Evolution \hspace*{10pt}
$\rlap{X}\square$ Cosmology and Fundamental Physics

\vspace{2mm}
\textcolor{myblue}{\textbf{\large Principal Authors}}\\
\begin{tabular}{ll}
Luke Zoltan Kelley & Maria Charisi \\
\textit{Northwestern University}    & \textit{California Institute of Technology} \\
LZKelley@northwestern.edu  \hspace*{50pt}  & mcharisi@caltech.edu
\end{tabular}

\vspace{2mm}
\noindent\textcolor{myblue}{\textbf{\large Co-authors}}\\
S. Burke-Spolaor,$^1$
J. Simon,$^2$
L. Blecha,$^3$
T. Bogdanovi\'c,$^4$
M. Colpi,$^5$
J. Comerford,$^6$
D. D'Orazio,$^7$
M. Dotti,$^5$
M. Eracleous,$^8$
M. Graham,$^9$
J. Greene,$^{10}$
Z. Haiman,$^{11}$
K. Holley- Bockelmann,$^{12,13}$ 
E. Kara,$^{14,15,16}$ 
B. Kelly,$^{14,15}$ 
S.~Komossa,$^{17}$ 
S. Larson,$^{18}$ 
X. Liu,$^{19}$ 
C.-P. Ma,$^{20}$ 
S. Noble,$^{15,21}$ 
V. Paschalidis,$^{22}$ 
R. Rafikov,$^{23}$ 
V. Ravi,$^{7,9}$
J. Runnoe,$^{24}$ 
A. Sesana,$^{25,5}$
D. Stern,$^2$
M. A. Strauss,$^{10}$
V. U,$^{26}$ 
M. Volonteri,$^{27}$ 
\& \mbox{the NANOGrav Collaboration}

\begin{adjustwidth}{0.5cm}{0.9cm}
\scriptsize{\it $^1$West Virginia University/Center for Gravitational Waves and Cosmology/CIFAR Azrieli Global Scholar;
$^2$Jet Propulsion Laboratory, California Institute of Technology;
$^3$University of Florida;
$^4$Georgia Institute of Technology;
$^5$Universita degli Studi di Milano-Bicocca;
$^6$University of Colorado Boulder;
$^7$Harvard-Smithsonian Center for Astrophysics;
$^8$The Pennsylvania State University;
$^9$California Institute of Technology;
$^{10}$Princeton University;
$^{11}$Columbia University;
$^{12}$Vanderbilt University;
$^{13}$Fisk University;
$^{14}$University of Maryland;
$^{15}$NASA Goddard Space Flight Center;
$^{16}$Massachusetts Institute of Technology;
$^{17}$Max Planck Institute for Radio Astronomy;
$^{18}$Northwestern University;
$^{19}$University of Illinois Urbana-Champaign;
$^{20}$University of California, Berkeley;
$^{21}$University of Tulsa;
$^{22}$University of Arizona;
$^{23}$University of Cambridge;
$^{24}$Univeristy of Michigan;
$^{25}$University of Birmingham;
$^{26}$University of California, Irvine;
$^{27}$Institut d'Astrophysique de Paris;
} \\
\end{adjustwidth}

\end{flushleft}

\vspace{-1mm}
\noindent \textcolor{myblue}{\textbf{\large Related White Papers}}

\small{
\noindent This is one of five core white papers written by members of the NANOGrav Collaboration.
\begin{enumerate}[topsep=0ex,itemsep=0.2ex,partopsep=0ex,parsep=0.2ex]
\item \textit{Nanohertz Gravitational Waves, Extreme Astrophysics, and Fundamental Physics\\ with Pulsar Timing Arrays}, J.~Cordes, M.~McLaughlin, et al.
\item \textit{Supermassive Black-hole Demographics \& Environments With Pulsar Timing Arrays}, \\ S.~R.~Taylor, S.~Burke-Spolaor, et al.
\item \textit{Physics Beyond the Standard Model with Pulsar Timing Arrays}, X.~Siemens, et al.
\item \textit{Fundamental Physics With Radio Millisecond Pulsars}, E.~Fonseca, et al.
\end{enumerate}
}

\vspace{2mm}
\noindent\textcolor{myblue}{\textbf{\large Abstract}} \\
\small{
    Pulsar timing arrays (PTAs) are on the verge of detecting low-frequency gravitational waves (GWs) from supermassive black hole binaries (SMBHBs).  With continued observations of a large sample of millisecond pulsars, PTAs will reach this major milestone within the next decade.  Already, SMBHB candidates are being identified by electromagnetic surveys in ever-increasing numbers; upcoming surveys will enhance our ability to detect and verify candidates, and will be instrumental in identifying the host galaxies of GW sources.  Multi-messenger (GW and electromagnetic) observations of SMBHBs will revolutionize our understanding of the co-evolution of SMBHs with their host galaxies, the dynamical interactions between binaries and their galactic environments, and the fundamental physics of accretion. Multi-messenger observations can also make SMBHBs `standard sirens' for cosmological distance measurements out to $z\simeq0.5$.  LIGO has already ushered in breakthrough insights in our knowledge of black holes.  The multi-messenger detection of SMBHBs with PTAs will be a breakthrough in the years 2020--2030 and beyond, and prepare us for LISA to help complete our views of black hole demographics and evolution at higher redshifts.
}

\setcounter{page}{0}
\setlength{\parindent}{4ex}
\thispagestyle{empty}

\newpage
\normalsize

\section{Multi-Messenger Science with Pulsar Timing Arrays}
\noindent Supermassive black holes (SMBHs) reside in the nuclei of massive galaxies \citep{Kormendy1995}.  Galaxy mergers deliver two SMBHs, along with massive inflows of gas, to the center of post-merger galaxies \citep{Begelman1980}.  Gravitationally bound SMBH binaries (SMBHBs) can then form, and eventually emit gravitational waves (GWs).  If sufficient gas remains, it can power bright active galactic nuclei (AGNs) \citep{Goulding2018,Koss2018}, observable across the electromagnetic (EM) spectrum.

In the coming decade, Pulsar Timing Arrays (PTAs) \citep{Hellings1983,Foster1990,Parkes2013,EPTA2013,McLaughlin2013}, like NANOGrav \citep{NANOGrav2015}, will likely detect GWs in the nano-Hertz frequency band, confirming the existence of SMBHBs \citep{Rosado2015,taylor2016,Kelley2018}. The expected signals are: (1) \textbf{Continuous Gravitational Waves} (CGWs) from individual, massive ($10^8$ -- $10^{10}\,{\rm M_{\odot}}$) and relatively nearby (redshifts $z \lesssim 0.5$) binaries, and (2) a stochastic \textbf{Gravitational Wave Background} (GWB) from the superposition of many unresolved SMBHBs \citep{Rajagopal1995,Phinney2001,Jaffe2003,Wyithe2003,Enoki2004,sesana2004,NG11yrGWB,NG11yrCW}.\footnote{It is unclear if the GWB will be detected first \citep{Rosado2015}, or both types of signals contemporaneously \citep{Mingarelli2017,Kelley2018}.}  At the same time, upcoming wide-field, time-domain (e.g., LSST), and multi-epoch spectroscopic (e.g., SDSS-V, DESI) surveys will discover an unprecedented number of SMBHB candidates.

As emphasized by the choice of thematic areas for Astro2020, and the ten NSF Big Ideas, the coming decade promises revolutions in multi-messenger astrophysics.  In this white paper, we discuss the astrophysics that is uniquely addressed with the detection of EM and nano-Hertz GW signals from SMBHBs.\footnote{See Holley-Bockelmann et al.\ for a discussion of mHz-GW science with higher-$z$, lower-mass sources.} In particular, we address four fundamental questions:

\begin{enumerate}[topsep=0ex,itemsep=0.2ex,partopsep=0ex,parsep=0.2ex,label=Q\arabic*.]

{\bfseries \item How do SMBHBs interact with their environments?}
SMBHBs evolve towards the GW regime through complex interactions with the galactic cores, e.g., stellar scatterings \citep{Merritt2005,Souza2017}, interactions with nuclear gas \citep{Fiacconi2013,Colpi2014}, and triple SMBH interactions from subsequent mergers \citep{Makino1994,Blaes2002,Hoffman2007,2010MNRAS.402.2308A}. The detection of the GWB will strongly constrain these physical processes, since each mechanism affects the shape of the GWB spectrum.

{\bfseries \item How does accretion in the presence of a SMBHB shape its EM signatures?}
Dynamical processes in the circumbinary disk induce a unique structure, which affects the resulting EM emission \citep{Schnittman2011,Dotti2012,spolaor2013}. However, AGNs with a single SMBH may mimic these signatures, making EM detections of sub-parsec binaries ambiguous \citep{Eracleous1997,Liu2016}. Multi-messenger detections of SMBHBs will illuminate the origin of the EM counterparts, allowing for direct comparisons against typical AGN.

{\bfseries \item How do SMBHs co-evolve with their host galaxies?}
SMBH--galaxy scaling laws, like the \msigma\ relation, indicate that SMBHs evolve symbiotically with their hosts \citep{Kormendy2013}.  PTA upper limits on the GWB already constrain these scalings \citep{Simon201603}, which will become more stringent with a GWB detection.  Mass measurements directly from CGWs will test and calibrate EM-based methods, while also assessing potential biases in SMBH--host scaling laws, which may be significant \citep{McConnell2013,Reines2015,shankar2016}.

{\bfseries \item How can binaries be used as cosmological probes?}
If the host galaxy of a SMBHB is identified, we can measure its redshift via spectroscopy and the luminosity distance from the GW signal, turning SMBHBs into standard sirens \citep{Schutz_1986,Holz2005}, as LIGO did with the detection of a NS merger \citep{Abbott2017}. The LISA mission \citep{amaro-seoane2012} can use massive binaries as standard sirens to even higher precisions and redshifts \citep{Kocsis2008,Tamanini2016,DelPozzo2018}.  PTAs will contribute independent siren measurements, will establish the procedure for LISA follow-up strategies, and tune rate predictions in the LISA band.

\end{enumerate}

\noindent PTAs already provide insights on SMBHB candidates.  For example, a binary proposed in the galaxy 3C\,66B was ruled out by the non-detection of GWs at the predicted amplitude and frequency \citep{Jenet2004}. Additionally, the GWB (see Fig.~1, C1) inferred from the population of quasars with periodic variability \citep{Graham2015,Charisi2016} is likely inconsistent with current PTA limits, indicating contamination with false detections \citep{Sesana201703}.  These examples speak to the tremendous potential of low-frequency GWs experiments for multi-messenger inference.

\section{Electromagnetic and Gravitational Wave Signals}

\begin{figure}
    \setlength{\abovecaptionskip}{2pt plus 10pt minus 4pt}
    \setlength{\belowcaptionskip}{-10pt}
    \centering
    \includegraphics[width=\textwidth]{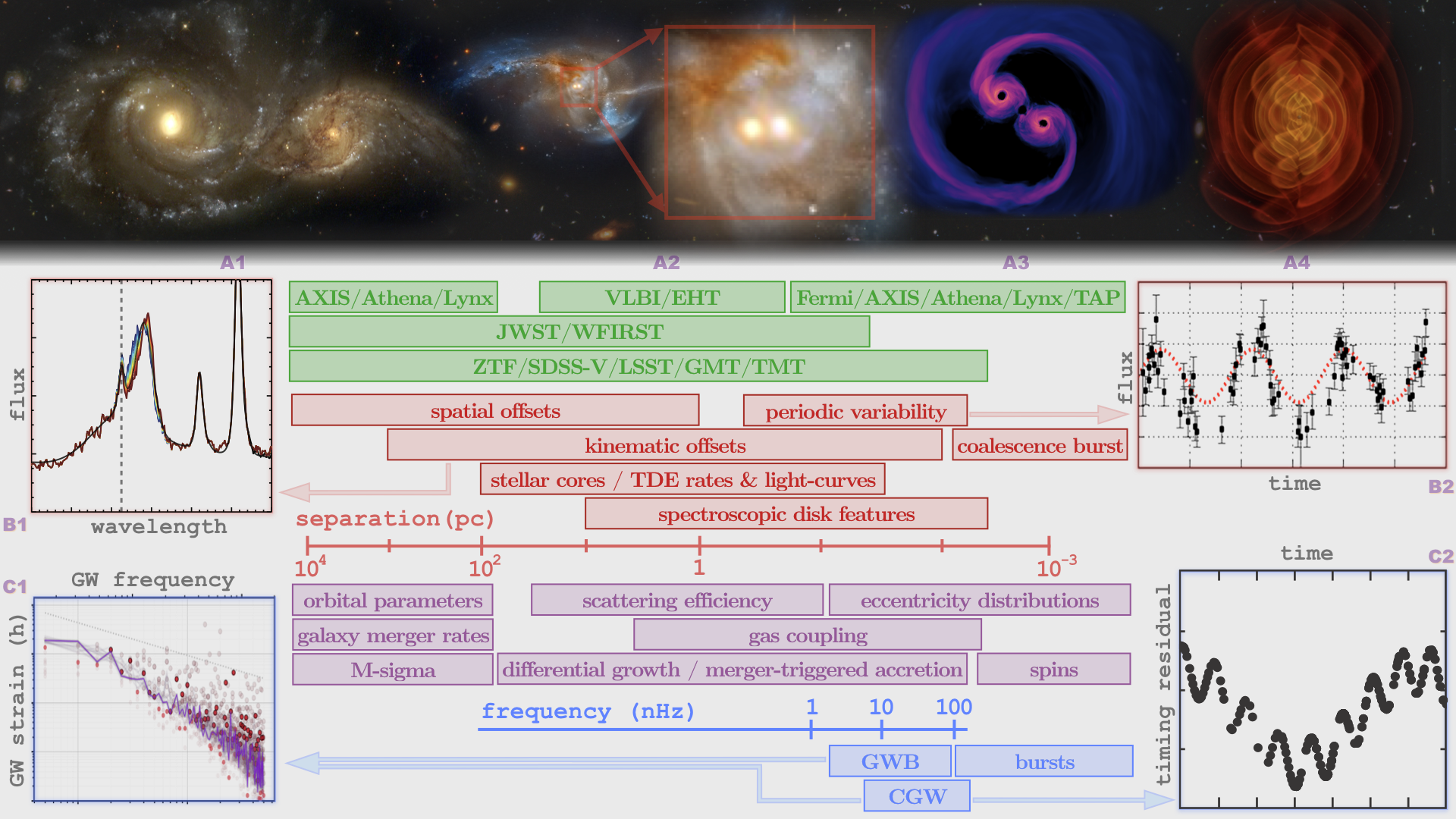}
\vspace{-4mm}
\caption{\small SMBHB evolution begins when galaxies merge (A1), progresses through environmental interactions with dark matter, stars, and gas (A2 \& A3), and ends with strong GW emission (A4).  EM signatures (red, also B1 \& B2) can be observed with current and future telescopes and surveys across the EM spectrum (green).  Combined with GW signals (blue \& C1, C2) from PTAs, the complex and uncertain physical processes (purple) can be tightly constrained. C1 shows the GWB and CGW sources from mock PTA observations.  C2 shows the theoretical timing residuals of a CGW source with high SNR.
    \textit{\footnotesize Subpanels: A1:The Hubble Heritage Team, A2: mockup with Hubble Legacy/M.~Pugh and SDSS; A3: \citep{2018ApJ...865..140D}; A4: NASA/C.~Henze;  B1 \citep{Runnoe2017}, B2 \citep{Graham2015}, C1 \citep{Kelley2018}, C2 \citep{Jenet2004}}.}
    \label{fig:1}
\vspace{-2mm}
\end{figure}

\noindent In Fig.~1, we show the expectations of how EM (red) and GW (blue) signals should follow the evolution of a SMBHB from large to small separations.  Because accretion onto both SMBHs can continue throughout the merger process \citep{Artymowicz1996,Shi2015}, numerous overlapping signatures provide the potential for robust multi-messenger and multi-wavelength discoveries.

\subsection{Formation and Evolution of SMBHBs}

Following a galaxy merger, SMBHs evolve to smaller separations in three main stages \citep{Begelman1980}:\\
\noindent\textbf{Large Separations} \emph{($\sim$kpc)}: Dynamical friction drags each SMBH, along with bound gas and stars, towards the center of the common potential \citep{Antonini2012,Pfister2017}.\\
\noindent\textbf{Intermediate Separations} \emph{($\sim$pc)}: Three-body interactions with stars in the galactic core \citep{merritt2013,Vasiliev2015}, and interactions with a nuclear disk \citep{Cuadra2009,Shi2012,Colpi2014,Fontecilla2019} extract energy from the binary\footnote{Though, at times, energy may be deposited \citep{Miranda2017,Munoz2019,Moody2019}.}.\\
\noindent\textbf{Small Separations} \emph{($\sim$mpc)}: The SMBHB decay is dominated by the emission of GWs.

The intermediate step remains the most uncertain\footnote{In the `\emph{final-parsec problem}', stellar scatterings may extract insufficient energy, but the growing consensus is that in general the `loss cone' (the stars able to interact with the binary) is sufficiently replenished \citep{Khan2013, Vasiliev2015}.}, since it involves complex interactions between the SMBHB and its local environment, which are challenging to tackle theoretically or resolve numerically. The same uncertainties affect LISA massive binary sources.  GWB measurements, or the detection of CGWs with associated EM measurements of their galactic environments, will provide crucial constrains on binary evolution and rates (\textbf{Q1,Q4}).

\subsection{Electromagnetic Signatures of SMBHBs}

\noindent
Observations of sub-parsec SMBHBs are very challenging, due both to their intrinsic rarity and extremely small angular separations. Gravitationally unbound dual AGN, at kpc separations, can be observed with high-resolution IR, optical, and X-ray instruments, and inform galaxy merger rates and initial orbital parameters \citep{Comerford2012,Liu2013}.  However, (sub-)parsec scale \textit{bound} systems can only be spatially resolved with radio Very Long Baseline Interferometry (VLBI). Indeed, VLBI has provided the record-holding binary at a projected separation of 7\,pc \citep{rodriguez2006} and, as of last year, an intriguing candidate at 0.35\,pc \citep{Kharb2017}. The exquisite resolution of current milli-meter VLBI, like the Event Horizon Telescope, holds tremendous promise for resolving sub-pc binaries, especially in the nearby universe \citep{Dorazio_2018}.

Once SMBHB separations fall below the resolution of even VLBI, we can still infer their presence by observing effects of the orbital motion or by identifying features indicative of an ongoing merger. Below we summarize some of these EM signatures (\textbf{Q2}):\\
\noindent\textbf{AGN with Doppler-shifted broad lines:}
Broad emission lines in AGN spectra arise from gas close to SMBHs, and may track their orbital motion (Fig.~1, B1) \citep{Nguyen2016,Nguyen2019}.  About a hundred candidate SMBHBs have been identified \citep{Tsalmantza2011,Bon2012,Eracleous_2012,Ju2013,Decarli2013,Shen2013,Liu2014a,McGurk2015,Runnoe2017,Guo2019}.\footnote{Double-peaked narrow lines have been used similarly as a tracer to select dual AGN.} However, gas in the accretion flows of single AGN can produce similar features, contaminating this population \citep{Liu2016}. Future surveys with multi-epoch spectroscopy, like the BH mapper/SDSS-V (scheduled for 2020 \citep{Kollmeier2017}), will reveal additional candidates and provide further tests of known candidates. \\
\noindent\textbf{AGN with periodic variability:}
Circumbinary disk simulations predict that SMBHBs can produce bright emission, periodically modulated at the orbital period or its harmonics \citep{MacFadyen2008,Roedig2011,Shi2012,Noble2012,Farris2012,farris201310, Dorazio2013,Gold2014a,Gold2014b,2015MNRAS.452.2540D,Charisi2015,Bowen2018}. This may be more pronounced in X-rays, since X-ray emission arises closer to the SMBHs \citep{Roedig2014,Shi2016,Ryan2017,Tang2018,2018ApJ...865..140D}. In addition to the well-known blazar OJ~287 \citep{Valtonen2008}, candidates have been identified in recent time-domain surveys (Fig.~1, B2) \citep{Graham2015, Charisi2016, Liu2016}. However, the intrinsic red-noise variability in AGN \citep{MacLeod2010}, combined with relatively short baselines, can lead to false detections \citep{Vaughan2016,Liu2016a,Sesana201703,2018MNRAS.481L..74H}. Extended multi-band and high-cadence data from surveys like ZTF \citep{ztf2019} and LSST \citep{lsst2002} will test current candidates and discover many new ones \citep{Kelley2019}.
\\
\noindent\textbf{Additional signatures:} Many additional features may signify the presence of a binary, such as helical radio jets \citep{Roos1993,Romero2000}; relativistic Doppler-boost \citep{Dorazio_2015,Charisi2018}; periodic self-lensing \citep{DOrazio2018}; double peaked, or extremely broad and oscillating FeK$\alpha$ lines \citep{Sesana2012,McKernan2013,McKernan2015}; UV/X-Ray deficits, from truncated circumbinary disks \citep{Tanaka2012, 2012ApJ...761...90G,Roedig2014}; enhanced rates of tidal-disruption events and features in their light-curves  \citep{Chen2009,Liu2009,Liu2014,Komossa2016}. These, along with indicators of recent galaxy mergers (e.g., recent star-formation bursts, tidal tails, or `cored' stellar density profiles \citep{Lauer2002}) can be used both for the search of CGW hosts, and for confirming EM binary candidates.

\subsection{Multi-messenger Observations}

The localization error of PTA detections will be large (typically, $\sim$100s deg$^2$ \citep{Sesana2010b,Zhu2016}), making host-galaxy determination challenging. However, PTAs will detect massive and relatively nearby binaries, which limits the number of potential hosts within the error volume. Also, since PTAs will identify binaries long before coalescence, if there is a bright counterpart, it will be long-lived (decades to millenia).  However, post-merger galaxies may be highly obscured \citep{Koss2018}. Thus, wide-field IR and X-ray telescopes, like WFIRST and TAP, will be essential in identifying obscured AGN within the error volume, whereas high angular-resolution instruments, like AXIS, Lynx and JWST, will provide more detailed follow-up.

Table 1 highlights important advances that are attainable \emph{only} through multi-messenger observations. Such detections reveal the interaction of SMBHBs with their galactic environments (\textbf{Q1}), and decipher the nature of accretion and EM emission around SMBHBs (\textbf{Q2}). They can also calibrate SMBH--host scaling relations.  These scalings may be substantially biased and are especially uncertain at higher masses, which PTAs naturally probe \citep{Bennert2011,Graham2013} (\textbf{Q3}). Additionally, the identification of the host galaxy allows the measurement of redshift, which, in some cases, can be compared with the luminosity distance measured independently from the GW signal (\textbf{Q4}),\footnote{In general, measurement of the GW-frequency evolution is required to break the degeneracy between chirp-mass and distance.  With LISA and LIGO, the `chirp' ($df/dt$) itself can be observed, but this is unlikely for PTAs.  Instead, with high signal-to-noise ratio CGW observations, the `pulsar term' can be recovered (visually apparent in Fig.~1, C2), allowing for measurement of the luminosity distance.}  making SMBHBs cosmological standard sirens \citep{Holz2005,Ellis2013}. PTAs can detect SMBHBs at larger distances than LIGO can detect NS mergers, and thus can complement LIGO observations to cover a wider range of redshifts.  Answers to \textbf{Q1}--\textbf{Q4} will also shape the predictions and preparations for multi-messenger science with LISA \citep{elisa2013}.

\newcommand{\rot}[2]{\rotatebox[origin=r]{90}{\,\textbf{Q#1:} \parbox[c]{2cm}{\centering#2}}}

\begin{table}[!ht]
\centering
\footnotesize
\begin{tabular}{p{0.04\textwidth}|p{0.28\textwidth}|p{0.28\textwidth}|p{0.3\textwidth}}
 & GW only & EM only & GW $+$ EM \\ \hline \hline

\rot{1}{SMBHB evolution} & \textit{GWB} spectrum (slope, amplitude, and features) constrain SMBHB/environment interactions.  \textit{CGW} detections will identify small-separation binaries in the GW regime.
&\textit{\mbox{X-ray}/Radio} surveys can directly image kpc binaries. \textit{mm/Radio VLBI} can spatially resolve parsec binaries. \textit{\mbox{X-ray}/Optical/IR/Radio} can identify mergers and track star/gas dynamics.  & \textbf{Directly maps host-galaxy properties to corresponding binary evolutionary stages. Provides a complete sequence of evolution from kpc to sub-pc scales} and associated timescales. \\ \hline

\rot{2}{Accretion \& Observables} & Detection of \textit{CGWs} will provide measurements of orbital frequency, eccentricity, phase, and a degenerate mass/distance. & EM signatures provide binary candidates.  Binaries are cross-checked with other indirect indicators, but may not conclusively confirm the binary nature of candidates. & Definitive detection of SMBHB with host galaxy \textbf{allows for in-depth multi-wavelength studies of binary emission.} With known binary parameters, measurements of e.g., Eddington ratio, radiative efficiency and spins are possible. \\ \hline

\rot{3}{SMBH--host co-evolution} & \textit{CGW} detection will provide a measurement of the chirp mass divided by the distance to the binary. Sources with high SNR or rapid frequency evolution yield independent distances. & \textit{\mbox{X-ray}/Optical/IR/UV} data will constrain the host galaxy properties (bulge mass, velocity dispersion, virial mass, Sersic index, etc), and potentially binary mass ratio. & \textbf{Directly measure SMBH masses and host-galaxy properties.  Calibrate scaling relations} without suffering from EM mass-measurement biases. \\ \hline

\rot{4}{Cosmological Distances} & \textit{CGWs} allow for independent measurement of luminosity distance, if GW frequency evolution is determined. & \textit{Optical/IR/UV} spectra will constrain or measure the host galaxy redshift. \textit{\mbox{X-ray}} can constrain AGN redshifts directly with iron line detections. & \textbf{Use SMBHBs as `standard sirens' to constrain cosmological parameters}, potentially at larger redshifts than accessible by LIGO, and with independent errors from the supernovae distance-ladder. \\ \hline

\end{tabular}
\caption{Summary of major advances attainable through GW and EM observations.}
\label{tbl:1}
\vspace{-6mm}
\end{table}

\section{Key Detectors \& Requirements}
\noindent
Here, we summarize particular key efforts required to realize these science opportunities.

\noindent \textbf{GW Observations}: Current PTAs are approaching the sensitivities required to detect GWs \citep{NG11yrGWB, NG11yrCW}. To ensure the discovery of GWs from SMBHBs, PTA collaborations need access to radio telescopes with large collecting areas operating in the frequency range of $\sim$100s MHz to a few GHz, such as large single-dish telescopes (like Arecibo and the GBT) or dish-arrays with equivalent sensitivities (like the proposed DSA-2000 and ngVLA). PTA collaborations, like NANOGrav, are long-timescale projects requiring continued monitoring of $\gtrsim 50$ pulsars with bi-weekly cadence over the coming years to decades and utilizing substantial amounts of telescope time ($\sim$ 1000s hours per year). Furthermore, the detection and characterization of GWs also necessitates the continuous development and improvement of statistical analysis techniques and infrastructures, in addition to finding new high-precision millisecond pulsars.

\noindent \textbf{EM Observations}: Upcoming time-domain and spectroscopic surveys will provide large samples of quasars. In order to efficiently distinguish binaries from AGNs with a single SMBH (i.e. minimize false detections), it is necessary to develop advanced statistical models of AGN variability (photometric and spectroscopic) over broad timescales.  Long-term and multi-wavelength monitoring is critical to exclude false positives, and access a sufficient parameter space of SMBHB orbital periods \citep{Kelley2019}. Ample access to smaller-scale telescopes for dedicated follow-up campaigns and searches for multiple binary tracers (e.g., shifted broad lines in photometric candidates) is also necessary. Complete catalogs of massive galaxies out to distances of a few Gpc will significantly aid host galaxy identification \citep{2019MNRAS.485..248G}.  Once host galaxies are identified, VLBI and thirty-meter class telescopes, like GMT and TMT, coupled with adaptive optics, can produce maps of the inner structures of the galaxies and provide detailed information of the SMBHB environments.

\noindent \textbf{Theory and computation}:
Improvements in the theoretical predictions of EM signatures from binary AGN will drastically improve our ability to detect SMBHB candidates, rule out false-positives, and eventually identify host galaxies following CGW detections.  Next-generation 3D simulations of circumbinary disks must be developed to include, for example, the effects of radiation and feedback, and follow binary evolution for more than a small number of orbits.  These disk-scale simulations must also be consistently coupled to realistic environments provided by cosmological simulations. Similarly, improvements in simulations of binary-stellar interactions are also required, in particular, expanding to realistic timescales and galactic environments.  Once GWs are detected, these models will be crucial in decoding binary parameters from GW+EM observations, both of CGW sources and the GWB spectrum.

\clearpage
\bibliographystyle{unsrtnat-truncauth-notitle}
\bibliography{refs.bib}

\end{document}